\documentclass[12pt, technote, twocolumn]{IEEEtran}

\usepackage{graphicx}           
\usepackage{wrapfig}            
\usepackage{verbatim}           
\usepackage{amsmath}            
\usepackage{amssymb}            
\usepackage{float}              
\usepackage{booktabs}           
\usepackage[usenames, 
            dvipsnames]{xcolor} 
\usepackage{bold-extra}
\usepackage{siunitx}            
\usepackage[affil-it]{authblk}  
\usepackage[noadjust]{cite}     
\usepackage{url}                
\usepackage{textcomp}

\DeclareSIUnit \parsec {pc}

\graphicspath{{./fig/}}

\newcommand{\nova}{NOvA}
\newcommand{\snews}{SNEWS}

\newcommand{\fig}{Fig.}

\begin{document}
  \title{\sc Observing the Next Galactic Supernova with the \nova{} Detectors \rm}

  \author[1]{Justin A. Vasel}
  \author[2]{Andrey Sheshukov}
  \author[3]{Alec Habig}
  
  \affil[1]{Indiana University \authorcr \emph{Email: jvasel@indiana.edu}
  \vspace{1.5ex}}
  \affil[2]{Joint Institute for Nuclear Research, Dubna \authorcr \emph{Email:
  sheshuk@nusun.jinr.ru} \vspace{1.5ex}}
  \affil[3]{University of Minnesota Duluth \authorcr \emph{Email:
  habig@neutrino.d.umn.edu} \vspace{1.5ex}}
  
  \onecolumn
  \maketitle
  \vspace{-3.5ex}
  \begin{center}
    on behalf of the \nova{} Collaboration \\
    \vspace{4ex}
    \footnotesize
    \emph{Talk presented at the APS Division of Particles and Fields Meeting
    (DPF 2017), July 31-August 4, 2017, Fermilab. C170731}
    \normalsize
  \end{center}
  
  \vspace{6ex}
  
  \begin{abstract}
    \noindent
    The next galactic core-collapse supernova will deliver a wealth of
    neutrinos which for the first time we are well-situated to measure. These
    explosions produce neutrinos with energies between 10 and
    \SI{100}{\mega\electronvolt} over a period of tens of seconds. Galactic
    supernovae are relatively rare events, occurring with  a frequency of just
    a few per century. It is therefore essential that all neutrino detectors
    capable of detecting these neutrinos are ready to trigger on this signal
    when it occurs. This poster describes a data-driven trigger which is
    designed to detect the neutrino signal from a galactic core-collapse
    supernova with the \nova{} detectors. The trigger analyzes
    \SI{5}{\milli\second} blocks of detector activity and applies background
    rejection algorithms to detect the signal time structure over the
    background. This background reduction is an essential part of the process,
    as the \nova{} detectors are designed to detect neutrinos from Fermilab's
    NuMI beam which have an average energy of \SI{2}{\giga\electronvolt}--well
    above the average energy of supernova neutrinos.
  \end{abstract}
  
  \vspace{1in}
  \twocolumn
  
  \section{Motivation}
  \noindent
  During a core-collapse supernova, only 1\% of the released gravitational
  binding energy is in the form of light and kinetic energy of ejecta. The other
  99\% is released in the form of neutrinos. The next galactic core-collapse
  supernova therefore represents an incredible opportunity for learning in
  several arenas of physics. The astrophysical implications include the ability
  to probe the conditions within the progenitor core throughout the explosion
  process. One aspect of the explosion that is of particular interest is the
  evolution of the shock front as it passes through the core before eventually
  blowing away the outer layers of the stellar envelope. Current simulations
  show that neutrinos play a vital role in driving the explosion by depositing
  heat behind the shock\cite{Janka2017}. Without this mechanism, the shock
  eventually stalls before crossing the boundary of the core and the explosion
  ultimately  fails to proceed.

  Supernova neutrinos also carry with them valuable information regarding the
  neutrinos themselves. Comparing the neutrino and photon arrival times can
  allow further constraints on the absolute neutrino mass, for example.
  Furthermore the spectral shape of the neutrino energy is dependent on the
  neutrino mass ordering and oscillation physics\cite{Scholberg2017}. In this
  way, supernova neutrinos will make a great compliment to the data provided by
  current accelerator- and reactor-based neutrino oscillation experiments, but
  they have the potential to provide us with insight that cannot be acquired
  with experiments on Earth. For example, the densities in the core are
  sufficiently dense that $\nu$--$\nu$ scattering interactions become relevant,
  a phenomenon that has yet to be observed.
  
  The insight provided by supernova neutrinos combined with the rarity of
  core-collapse supernovae in a galaxy like ours ($3 \pm 1$ per
  century\cite{Cappellaro2001}) underscores the importance of being prepared to
  record the data so that the opportunity is not missed. The \nova{} experiment
  is making preparations to that end and will be well-situated to make a
  substantive measurement of the supernova neutrino flux when it arrives. 
  
  \section{The \nova{} Detectors}  
  \noindent
  \nova{} is a long-baseline neutrino oscillation experiment which makes
  precision measurements of the $\nu_\mu \rightarrow \nu_e$ and $\bar{\nu}_\mu
  \rightarrow \bar{\nu}_e$ oscillation probabilities. Through these
  measurements, \nova{} is setting competitive limits on mass ordering, the
  CP-violating phase $\delta_{\text{CP}}$, and a determination of the
  $\theta_{23}$ octant. 
  
  The \nova{} detectors are composed of extruded PVC cells which are filled with
  liquid scintillator. A wavelength-shifting fiber is looped through each cell
  and connected to an avalanche photo diode which serves to collect the light
  produced by particle interactions in the scintillator. The near detector at
  Fermilab is \SI{300}{\tonne} and is situated \SI{105}{\metre} underground. The
  far detector weighs \SI{14}{\kilo\tonne} and sits \SI{810}{\kilo\metre} away
  from Fermilab in Ash River, MN at the Earth's surface. There is a modest
  barite overburden at the far detector to provide some shielding from cosmic
  rays.
  
  \begin{figure}[ht!]
      \includegraphics[width=\columnwidth]{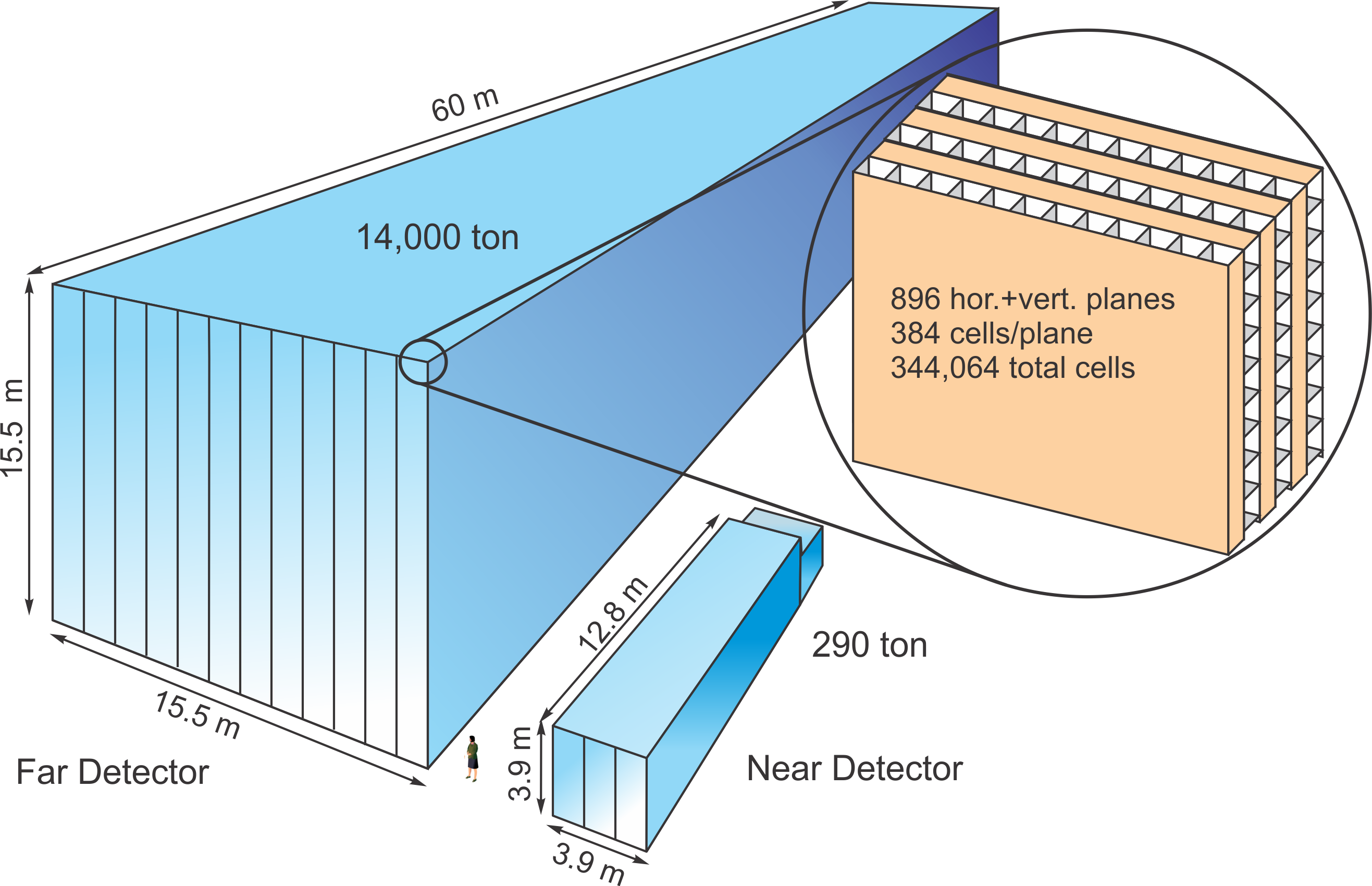}
      \caption{The \nova{} detectors. A \SI{300}{\tonne} near detector dwells
      \SI{100}{\metre} underground. The \SI{14}{\kilo\tonne} far detector sits 
      on the surface, with a modest barite overburden.}
      \label{fig:dets}
  \end{figure}

  \section{Supernova Neutrinos}
  \noindent
  The supernova explosion process plays out in four stages: collapse, bounce,
  accretion, and cooling. In the collapse phase, nuclear fusion has ceased in
  the iron core and it begins to collapse under its own weight. As the collapse
  proceeds, the $\nu_e$ flux increases sharply due to electron capture on nuclei
  and free  protons, a process known as ``neutronizaton''. \begin{equation}
  e^{-} + p \rightarrow \nu_e + n \end{equation} As the collapse proceeds, the
  density grows. Eventually, densities become large enough that the neutrino
  diffusion time becomes significant and the neutrinos become trapped. The
  result is the formation of a ``neutrinosphere'', a surface analogous to a the
  photosphere of a star. At this point, the neutrino luminosity begins to
  decrease.
  
  When the collapsing core reaches nuclear densities ($\rho \gtrsim$
  \SI{3e14}{\gram\per\cubic\centi\metre}), the strong nuclear force becomes
  repulsive and the core becomes incompressible, causing the in-falling matter
  to rebound. This is the bounce phase of the supernova. The result of this
  bounce is a shock wave that propagates out towards the surface of the iron
  core. Once the shock front passes the neutrinosphere, the neutrinos are free
  to escape the star. This burst of neutrinos is observable in the flux and at
  this time the neutrino luminosity is the largest it will be at any point
  during the explosion.

  The shock wave pushes against in-falling matter as it travels outward.
  Additionally, photons produced in the shock region are sufficiently energetic
  to dissociate iron nuclei, an endothermic interaction which robs energy from
  the shock front. Modern simulations show that the shock front stalls before
  breaking out of the core and is destined to eventually fall back inward; the
  explosion fails. But we know from observation that stars successfully explode
  all the time. One promising solution to this discrepancy is that neutrinos
  flowing outward and passing the shock front deposit a fraction of their energy
  into it and eventually revive the stalled shock, allowing the explosion to
  proceed. This is the accretion phase and is characterized by turbulent motions
  within the shock front which manifest as correspondingly complex features in
  the flux profile.

  After the shock front has reached the edge of iron core, blowing apart the
  stellar envelope, only a neutron star remains. This is the Kelvin-Hemholtz
  cooling phase, wherein the proto-neutron star releases its remaining
  gravitational binding energy in the form of $\nu$ and $\bar{\nu}$ of all
  flavors.
  
  This entire processes plays out in and instant as far as the late star is
  concerned. The neutronization burst during the collapse and bounce phases
  occurs within the first \SI{100}{\milli\second} of the supernova. The
  accretion phase lasts for the next \SIrange{800}{900}{\milli\second}. The
  final stage, cooling, starts nearly \SI{1}{\second} after the bounce and lasts
  for 10s of  seconds.

  \section{Detector Simulation}
\begin{figure}[ht!]
      \includegraphics[width=\columnwidth]{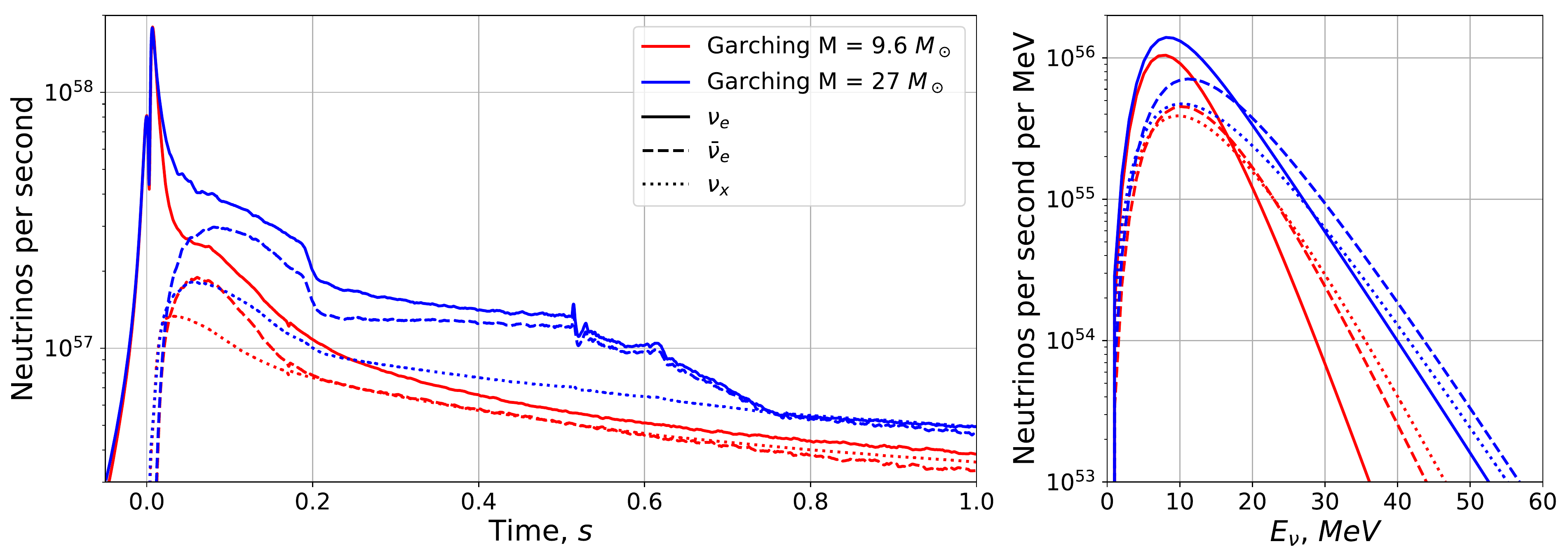}
      \includegraphics[width=\columnwidth]{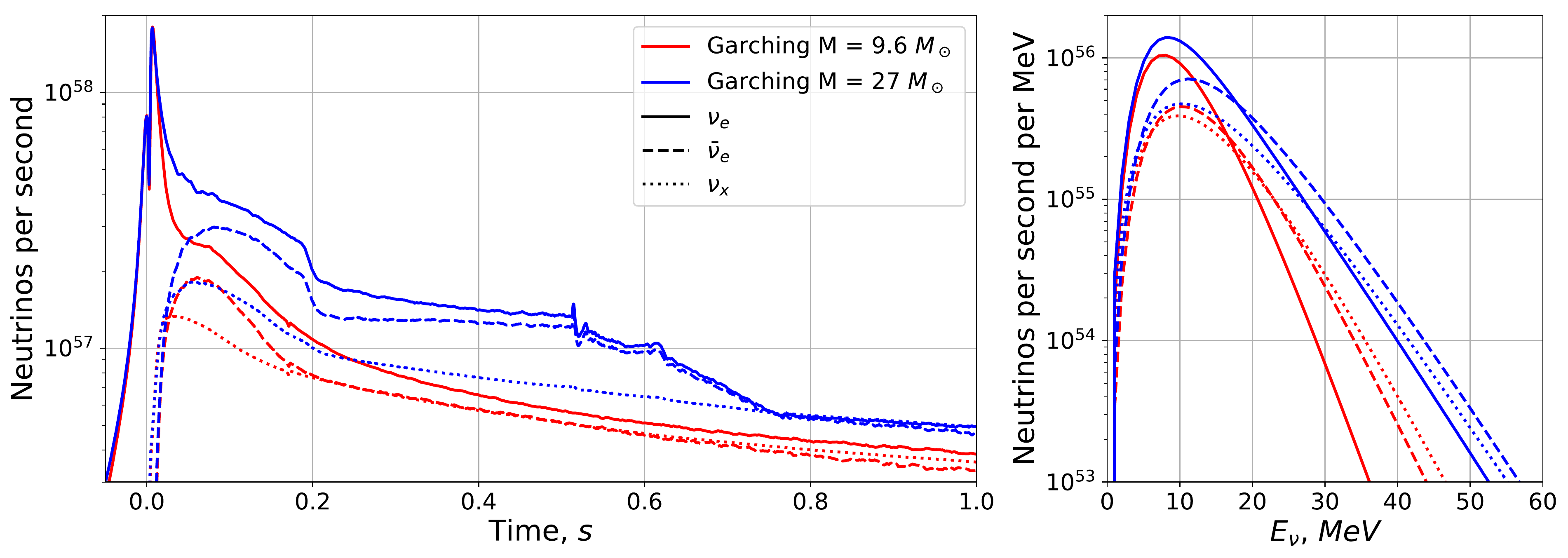}
      \caption{Simulated supernova neutrino flux time profile (top) and energy
      profile (bottom) based on the Garching\cite{garching} simulation for two
      progenitor masses.}
      \label{fig:flux}
  \end{figure}

  \noindent
  To simulate the supernova signal, the existing simulation framework at \nova{}
  is used with some modifications. This framework uses GENIE\cite{genie1,
  genie2} for particle generation and GEANT4\cite{geant4} for particle
  propagation within the detector geometries. To simulate supernovae, two
  modifications are needed: neutrino cross sections for energies in the MeV
  regime must be included in place of those in the GeV typical of our
  accelerator-based neutrinos, and neutrino fluxes for a simulated supernova.
  The fluxes we use are from the Garching group\cite{garching} and include
  several progenitor masses (\fig{} \ref{fig:flux}).
  
\begin{figure}[ht!]
      \includegraphics[width=\columnwidth]{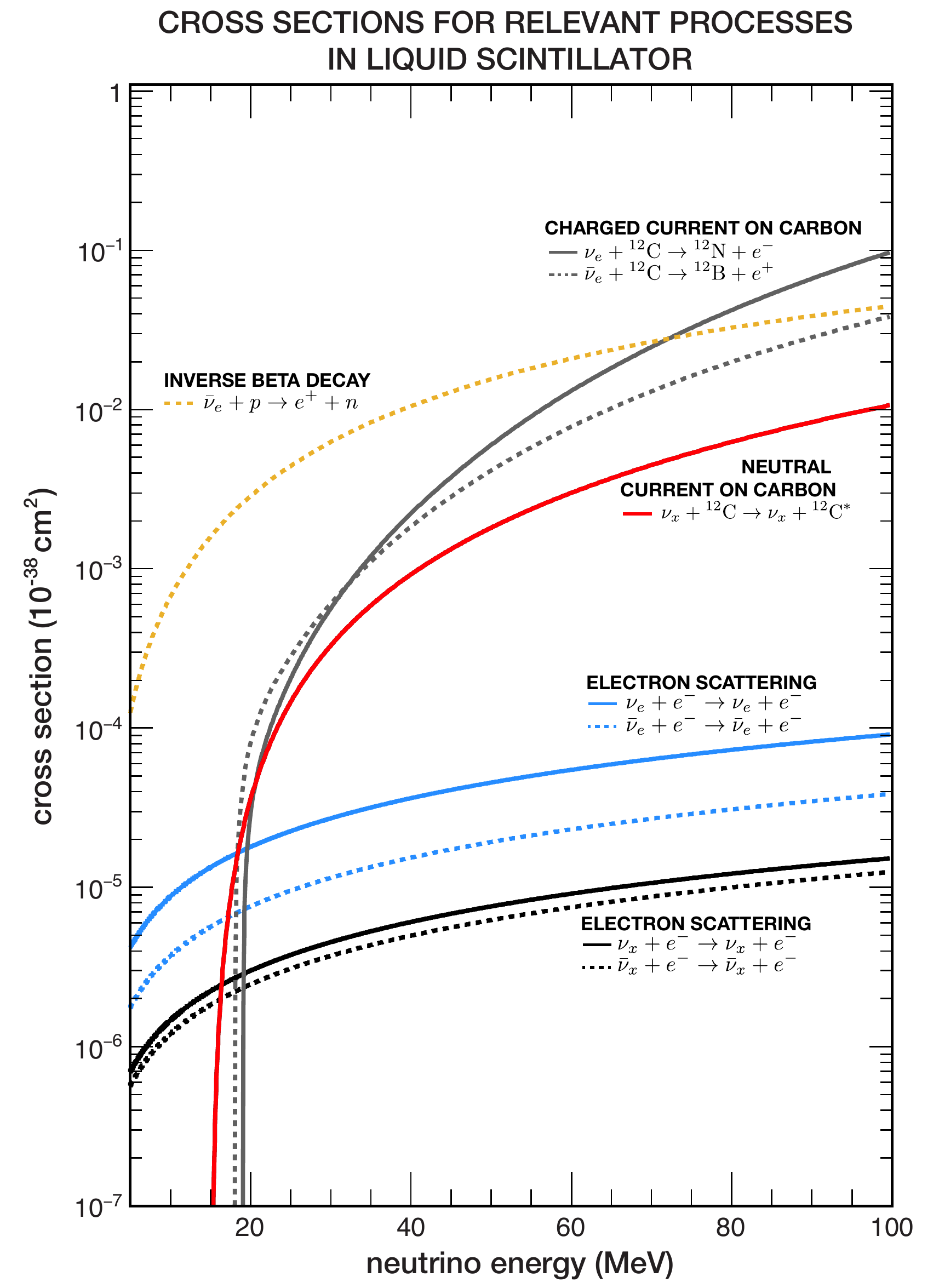} \caption{Cross
      sections for relevant processes in liquid scintillator.  Inverse beta
      decay is the dominant process at supernova neutrino energies. These cross
      sections were ported into our framework from the
      SNOwGLoBES\cite{snowglobes}  software, which collected and parsed them
      from a variety of sources.} \label{fig:xscns}
  \end{figure}

  The \nova{} detectors consist mainly of plastic PVC extrusions and liquid
  scintillator. Hydrogen and carbon are the most abundant nuclei in the
  detector. The most common neutrino interactions in the supernova energy regime
  are inverse beta decay (IBD) on free protons, charged and neutral current
  exchanges on ${}^{12}$C, and $\nu$--$e$ scattering (\fig{} \ref{fig:xscns}).
  
\begin{figure}[ht!]
      \includegraphics[width=\columnwidth]{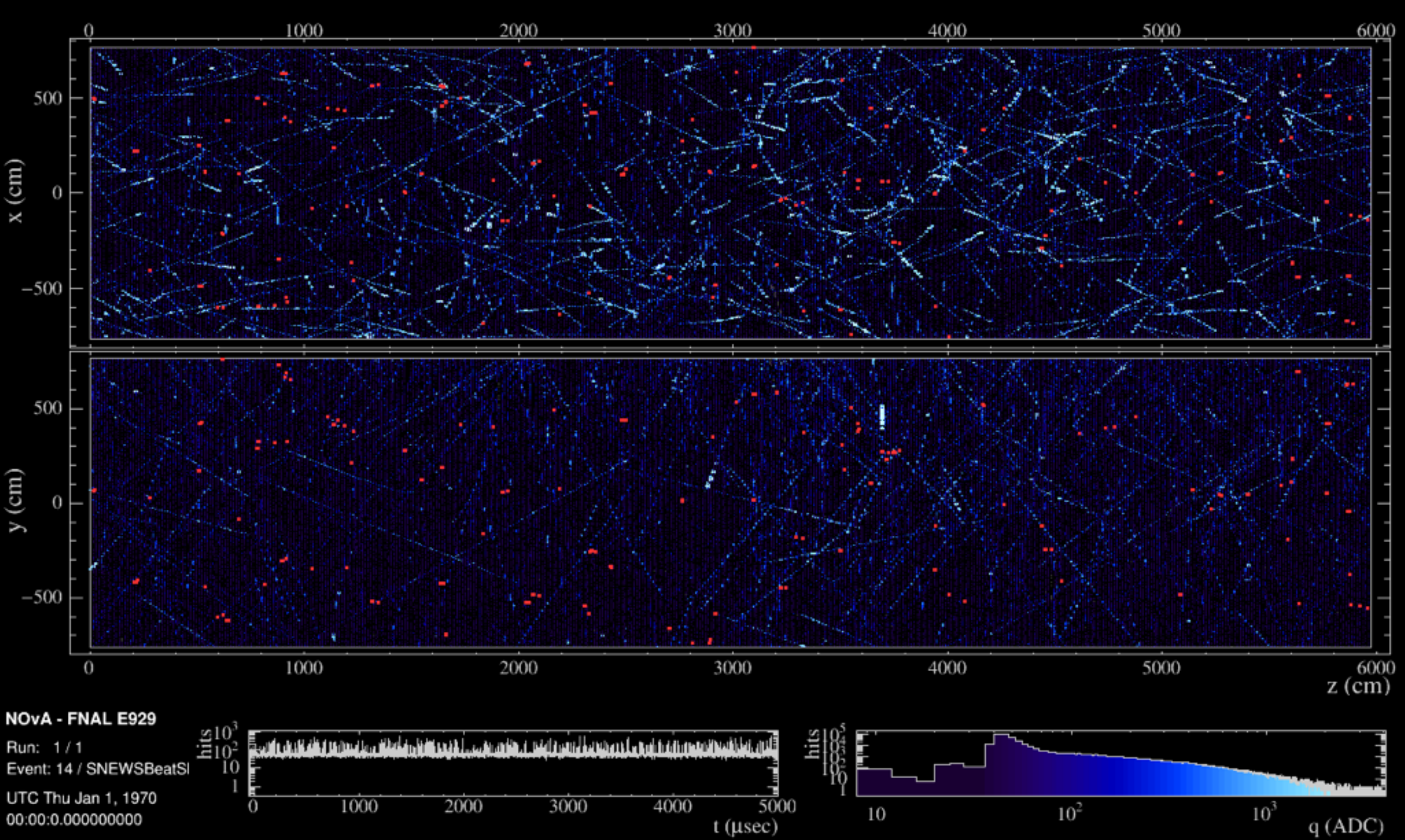}
      \caption{Event display of a simulated \SI{10}{\kilo\parsec} supernova over
      a period of \SI{5}{\milli\second}. Only IBD positrons are present in the
      signal in this example (red dots indicate IBD production locations). The
      blue hits are minimum-bias cosmic data which has been overlayed with the
      signal. This represents \SI{5}{\milli\second} of detector activity
      occurring 70ms after stellar core bounce.}
      \label{fig:evd2d}
  \end{figure}

  After the simulated particles have been generated and propagated, the detector
  readout chain is simulated to produce an observed signal which accounts for
  detector thresholding effects, smearing, and efficiencies. The final step in
  this simulation chain involves overlaying the simulated signal with real
  minimum-bias data. The result is a realistic representation of what we might
  expect to see in the event of a real galactic supernova. \fig{}
  \ref{fig:evd2d} is an example event display in the \nova{} far detector of the
  overlaid signal and minimum-bias data.
  
  \section{Triggering on Supernovae}
  \noindent
  The \nova{} experiment subscribes to the Supernova Early Warning System
  (\snews)\cite{snews}, a global network for existing neutrino detectors to
  disseminate news of a supernova neutrino signal when it arrives. To keep their
  false alarm rate low, \snews{} requires two independent experiments to report
  a supernova candidate within a \SI{10}{\second} coincidence window. When the
  alert arrives, both detectors at \nova{} will automatically trigger a long
  readout, writing a continuous stream of data to disk for as long as the DAQ
  can handle it. Currently \nova{}'s DAQ is capable of reading out
  \SIrange{45}{60}{\second}  of continuous data to disk. While \nova{}
  subscribes to alerts from \snews, it does not yet contribute to the network
  itself, but an internal trigger has been developed and is undergoing testing.
  \nova{} expects to join the \snews{} community in the near future.
  
  The current iteration of our developmental internal trigger relies on tagging
  interaction candidates from the dominant IBD channel. The occurrence rate of
  these candidates is monitored in real time. If this rate exceeds a chosen
  threshold, then the trigger fires and a long readout of both \nova{} detectors
  is executed. 

  The background rate at the far detector is much larger than that of the
  signal, so some selection is necessary to identify supernova signal
  candidates. The selection algorithm currently in use is the following:
  \begin{enumerate}
      \item \textbf{Remove hits from other identified physics} \\
      For instance, cosmic tracks and associated activity like Michel electrons.
      \item \textbf{Cluster hits in space and time} \\
      Reduce the chance we will accidentally tag noise activity, which registers
      an ADC value similar to our signal.  Detector noise is not spatially or
      temporally correlated.
      \item \textbf{Require clusters to have both an $x$-- and $y$--component} \\ 
      Having both views enables us to infer a 3D position in the detector and 
      allows for more accurate calibration. This further aids in reducing the 
      noise rate.
      \item \textbf{Cut on fiducial volume}
      \item \textbf{Cut on total cluster ADC}
  \end{enumerate}
  
\begin{figure}[ht!]
      \includegraphics[width=\columnwidth]{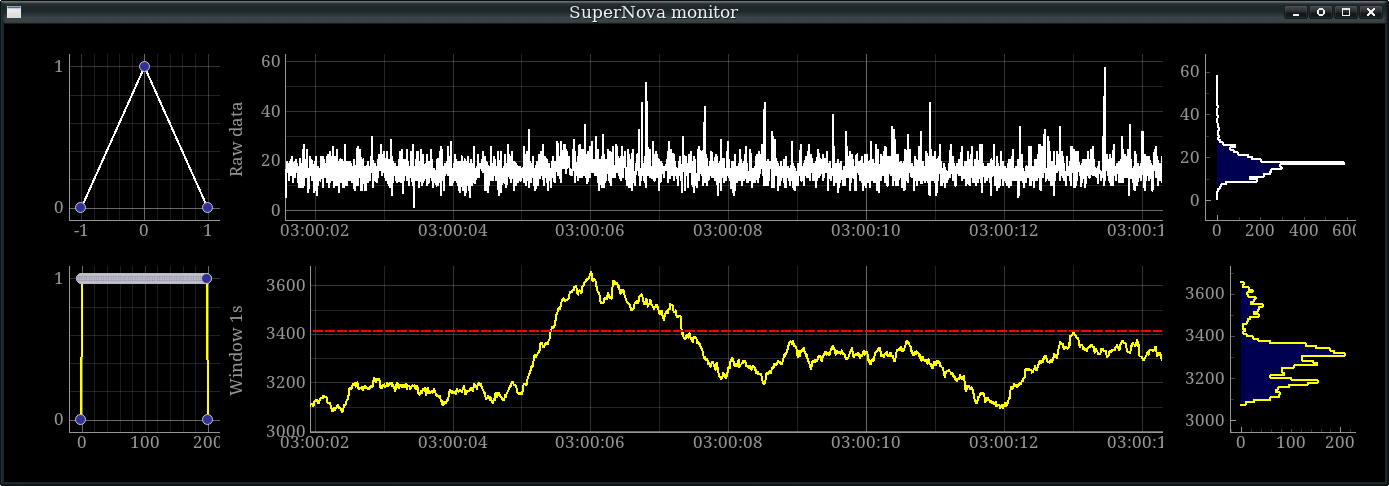} \caption{Time series
      analysis of supernova hit candidates over a period of \SI{12}{\second} for
      a simulated \SI{10}{\kilo\parsec} supernova overlaid with minimum-bias
      data. The top pane shows the raw time series, while the bottom pane shows
      the same data convolved with a \SI{1}{\second} summation kernel. The
      trigger will execute when this convolved signal exceeds a predefined
      threshold. In the bottom pane, this threshold is indicated by the dashed
      red line. This threshold can be adjusted to achieve an acceptable
      false-positive trigger rate.}
      \label{fig:snmonitor}
  \end{figure}

  Any hits that remain after applying these selection cuts are considered
  supernova hit candidates. \fig{} \ref{fig:snmonitor} shows two time series of
  the number of supernova hit candidates. A kernel is applied to the raw time
  series which sums the number of candidates over a sliding \SI{1}{\second}
  window, producing a much cleaner signal. If this enhanced signal exceeds a
  predetermined threshold, then the trigger will execute. When this occurs, both
  detectors begin to write continuous chunks of data to disk from their data
  buffers. They do this until the requested data in the buffer is exhausted to
  capture as much of the supernova signal as possible.
  
  In addition to capturing this data, once the trigger is fully tested and
  deployed it will also send an alert to \snews{}. In order to keep the overall
  false-alarm rate low, \snews{} requires participating experiments to limit
  their own false-positive rate to less than one per week. This requirement
  motivates our determination of the triggering threshold.
  
  \section{Sensitivity and Outlook}
\begin{figure}[ht!]
      \includegraphics[width=\columnwidth]{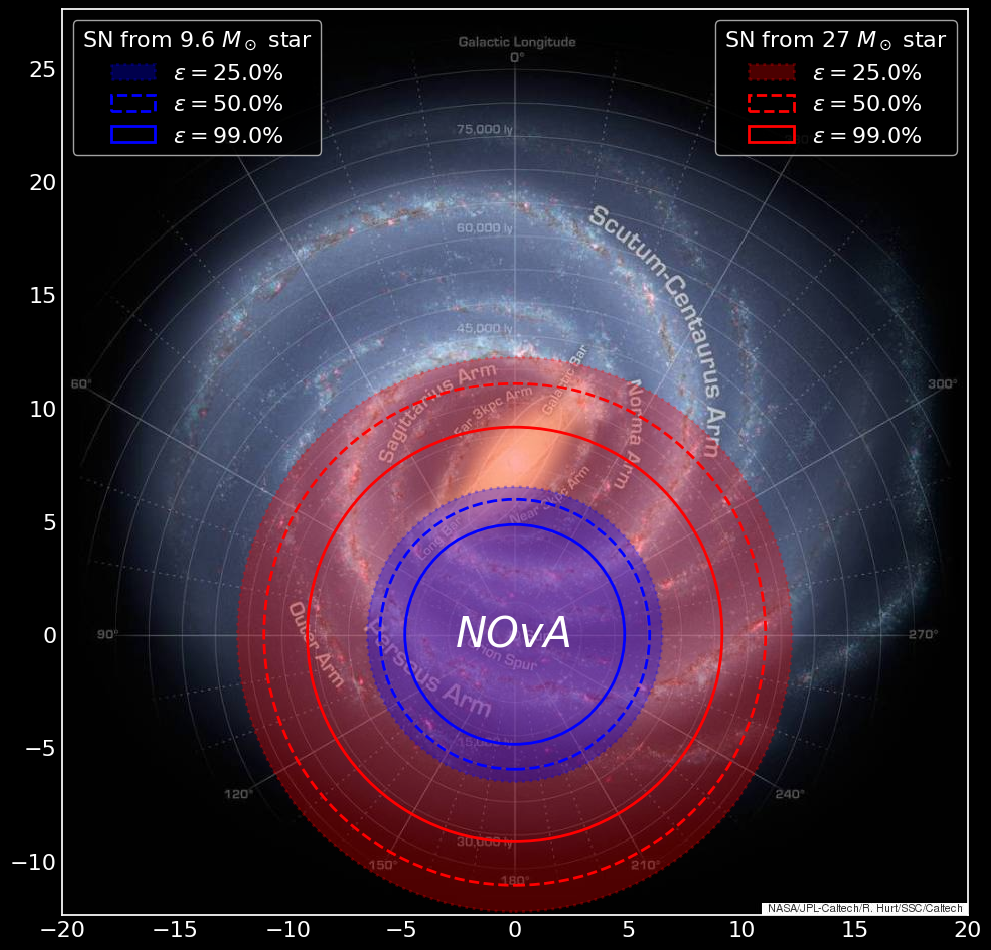}
      \caption{Current trigger efficiency contours mapped onto the Milky Way 
      illustrate \nova's sensitivity to galactic core-collapse supernovae for 
      progenitor stars of 9.6$M_\odot$ (blue) and 27$M_\odot$ (red).}
      \label{fig:galaxy}
  \end{figure}

  \noindent
  Based on the selection algorithm used to identify supernova hit candidates and
  the triggering thresholds we've chosen, our triggering efficiency for a
  supernova from a 27$M_\odot$ progenitor is around 99\% for supernova between
  Earth and galactic center. \fig{} \ref{fig:galaxy} shows efficiency contours
  overlaid on a map of the Milky Way. Ideally we would want to be sensitive to
  supernovae from anywhere in our galaxy and efforts are currently underway to
  achieve that. Improving the candidate selection efficiency and applying more
  sophisticated kernels on the time series data are being explored, as well as
  utilizing new technologies like machine learning and computer vision. 
  
  \vspace{-0.3in}
  
  \bibliographystyle{utphys}
  \bibliography{vasel-dpf-2017-proceeding}
  
\end{document}